# Conditional entropy; an alternative derivation

# of the pair correlation function for simple classical fluids

by Richard BONNEVILLE, emeritus

Centre National d'Etudes Spatiales (CNES), 2 place Maurice Quentin, 75001 Paris, France

phone: +33 6 87 60 78 41, mailto:richard.bonneville@outlook.fr

**Abstract**

We present an alternative derivation of the pair correlation function for simple classical fluids by using a variational approach. That approach involves the conditional probability $p(3,...,N/1,2)$ of an undefined system of N particles with respect to a given pair (1,2), and the definition of a conditional entropy $\sigma(3,...,N/1,2)$. An additivity assumption of $\sigma(3,...,N/1,2)$ together with a superposition assumption for $p(3/1,2)$ allows deriving the pair probability $p(1,2)$. We then focus onto the case of simple classical fluids, which leads to an integral, non-linear equation that formally allows computing the pair correlation function $g(R)$. That equation admits the one resulting from the hyper netted chain approximation (and the Percus-Yevick approximation) as a limit case.

## 1. Introduction

The knowledge of the pair correlation function $g(R)$ is of primary importance in the physics of classical fluids since it is connected to the equation of state and to the thermodynamical quantities [1]; it is also a key function in the interpretation of the X-ray scattering patterns, which allows going backward up to the molecular interactions [2]. For a given intermolecular potential, $g(R)$ can be determined by numerical simulations or by theoretical approaches. In the latter case, the ab initio process leads to an infinite hierarchy of equations, which can be approximately solved by assuming at some stage a truncation or a closure relation. Kirkwood's superposition approximation [3,4,5] is a well-known example of such a closure relation. More sophisticated approaches have been elaborated such as the hyper netted chain approximation [6,7] or the Percus-Yevick approximation [8,9], both being based upon the assumption of an additional relation allowing to solve the Ornstein–Zernike equation [10].

In the present paper we give an alternative derivation of the pair correlation function for simple classical fluids by using a variational approach. That approach involves the conditional probability $p(3,...,N/1,2)$ of an undefined system of N particles with respect to a given pair $(1,2)$, and the definition of a conditional entropy $\sigma(3,...,N/1,2)$. In order to introduce the conditional entropy concept and to become more familiar with it, we will first present in the next section the simpler case of the one particle probability $p(1)$. The free energy $F = \langle H \rangle - TS$ is minimized through a functional derivation with respect to $p(1)$; we introduce the conditional entropy $\sigma(2,3,...,N/1)$ and make an additivity assumption which allows going beyond the simple mean-field approximation. In the following section we will extend that approach to the derivation of the pair probability $p(1,2)$; the free energy is minimized with respect to $p(1,2)$ and we introduce the conditional entropy $\sigma(3,...,N/1,2)$. An additivity assumption of $\sigma(3,...,N/1,2)$ together with a superposition assumption for $p(3/1,2)$ allows deriving the pair probability. We will finally focus onto the case of simple classical fluids; we will derive an integral, non-linear equation which formally allows computing the pair correlation function $g(R)$.

## 2. One particle probability

Assuming pair interactions only, the system of N identical elements is described with obvious notations by the Hamiltonian

$$H = \sum_i h(i) + \sum_{i>j} U(i,j) \tag{1}$$

where $U(i,j) = U(j,i)$. We have $\sum_{\varphi_i,\varphi_j} p(i,j) = 1$, $p(i) = \sum_{\varphi_j} p(i,j)$, $\sum_{\varphi_i} p(i) = 1$.

The mean energy $\langle H \rangle$ and the statistical entropy S of the system are respectively given by

$$\langle H \rangle = \sum_i \sum_{\varphi_i} p(i) h(i) + \sum_{i>j} \sum_{\varphi_i,\varphi_j} p(i,j) U(i,j) \tag{2}$$

$$S = -k_B \sum_{\varphi_1,\varphi_2,...,\varphi_N} p(1,2,...,N) Log(p(1,2,...,N)) \tag{3}$$

where $p(1,2,...,N)$ is the probability law for the set of N elements; the sum runs over all the states $\varphi_i$ of the element i and over all the elements.

In this section we recall the process for obtaining the one particle probability $p(i)$ by minimizing the free energy $F = \langle H \rangle - TS$ through a functional derivation with respect to $p(i)$, i.e. by writing

$$\frac{d}{dp(i)}\left(F + \alpha \sum_i \sum_{\varphi_i} p(i)\right) = 0 \tag{4}$$

In equ.(4) the Lagrange parameter $\alpha$ expresses the normalization condition of the probabilities.

We can in a fully general way express $p(1,2,...,N)$ as $p(1)p(2,...,N/1)$ where $p(2,...,N/1)$ is the conditional probability law of 2,3,…,N with respect to 1; we have $\sum_{\varphi_2,...,\varphi_N} p(2,...,N/1) = 1$.

We can thus write

$$S = -k_B p(1) Log(p(1)) - k_B p(1) \sum_{\varphi_2,...,\varphi_N} p(2,...,N/1) Log(p(2,...,N/1)) \tag{5}$$

Consequently

$$\frac{dS}{dp(1)} = -k_B \sum_{\varphi_2,...,\varphi_N} p(2,...,N/1) \times Log(p(2,...,N/1)) - k_B Log(p(1)) - k_B \tag{6}$$

Besides, since $p(i,j) = p(i)p(j/i)$ with $\sum_{\varphi_j} p(j/1) = 1$, we have

$$\frac{d\langle H\rangle}{dp(1)} = h(1) + \sum_{j\neq 1}\sum_{\varphi_j} p(j/1)U(1,j) \tag{7}$$

We now introduce the conditional entropy

$$\sigma(2,...,N/1) = -k_B \sum_{\varphi_2,...,\varphi_N} p(2,...,N/1) Log(p(2,...,N/1)) \tag{8}$$

Minimizing the free energy then leads to

$$-k_B T Log(p(1)) = h(1) + \sum_{j\neq 1}\sum_{\varphi_j} p(j/1)U(1,j) - T\sigma(2,...,N/1) + \alpha' \tag{9}$$

The simple mean-field approximation neglects all correlation between the particles i.e. $p(j/1) \cong p(j)$ and $p(2,...,N/1) \cong \prod_{i=2}^{N} p(i)$; then the term $\sigma(2,...,N/1)$ is independent of 1, i.e. it is a constant, so that

$$p(1) \propto exp\frac{-1}{k_B T}\left(h(1) + \sum_{j\neq 1}\sum_{\varphi_j} p(j)U(1,j)\right) \tag{10}$$

But it is possible to go further by making for the conditional entropy $\sigma(2,...,N/1)$ an additivity assumption which is the following

$$\sigma(2,...,N/1) \cong \sum_{\varphi_i} \sigma(i/1) = -k_B \sum_{\varphi_i} p(i/1) Log(p(i/1)) \tag{11}$$

Equ.(9) is then changed into

$$h(1) + k_B T Log(p(1)) = -\sum_{j\neq 1}\sum_{\varphi_j} p(j/1)\left(U(1,j) + k_B T Log(p(j/1))\right) + \alpha' \tag{12}$$

The above equation equ.(12) and the consistency relation

$$p(j) = \sum_{\varphi_1} p(j/1) \tag{13}$$

together with the normalization conditions $\sum_{\varphi_i} p(i) = 1$ and $\sum_{\varphi_j} p(j/1) = 1$ allow to determine $p(1)$ and $p(j/1)$. In the case of an Ising-type system, that system of equations leads to results similar to those resulting from the Bethe-Peierls approximation [11].

### 3. Pair probability

We now try to evaluate the two-particle probability. For the sake of simplicity we assume that the Hamiltonian contributions $h(i)$ and $U(i, j)$ do not act upon the same dynamical variables of the particle system. For instance in a classical fluid $h(i)$ involves the translation motions and the internal degrees of freedom (rotation and vibration, coupling to an external field) whereas $U(i, j)$ accounts for molecular interactions only depending upon their relative position and orientation. After that separation of the variables, we henceforth can concentrate ourselves on the interaction terms. The following development would not apply to a situation such as an Ising-type system where the same dynamical variables appear in $h(i)$ and $U(i, j)$.

We will minimize the free energy by performing a functional derivation with respect to $p(1,2)$. We extend the approach of the previous section by writing $p(1,2,...,N) = p(1,2)p(3,...,N/1,2)$ where $p(3,...,N/1,2)$ is the conditional probability law of 3,…,N with respect to the pair (1,2); we have $\sum_{\varphi_3,...,\varphi_N} p(3,...,N/1,2) = 1$.

The energy term of the free energy is

$$\langle H \rangle = \sum_{i>j} \sum_{\varphi_i,\varphi_j} p(i,j) U(i,j) = \sum_{i>j} \sum_{\varphi_i,\varphi_j} p(1,2,...,N) U(i,j) \tag{14}$$

i.e.

$$\begin{aligned}
<H_{12}> &= \sum_{\varphi_1,\varphi_2} p(1,2) U_{12} \\
<H_{13}> &= \sum_{\varphi_1,\varphi_3} p(1,3) U_{13} = \sum_{\varphi_1,\varphi_2,\varphi_3} p(1,2) p(3,...N/1,2) U_{13} \\
<H_{23}> &= \sum_{\varphi_2,\varphi_3} p(2,3) U_{23} = \sum_{\varphi_1,\varphi_2,\varphi_3} p(1,2) p(3,...N/1,2) U_{23}
\end{aligned} \tag{15}$$

Consequently

$$\begin{aligned}
\frac{d\langle H \rangle}{dp(1,2)} &= U(1,2) + N\sum_{\varphi_3} p(3,...N/1,2) U(1,3) + N\sum_{\varphi_3} p(3,...N/1,2) U(2,3) \\
&= U(1,2) + N\sum_{\varphi_3} p(3/1,2) U(1,3) + N\sum_{\varphi_3} p(3/1,2) U(2,3)
\end{aligned} \tag{16}$$

As for the entropy term we can express it as

$$S = -k_B p(1,2) Log(p(1,2)) - k_B p(1,2) \sum_{\varphi_3,...,\varphi_N} p(3,...,N/1,2) Log(p(3,...,N/1,2)) \tag{17}$$

Consequently

$$\frac{dS}{dp(1,2)} = -k_B \left( 1 + Log(p(1,2)) + \sum_{\varphi_3,...,\varphi_N} p(3,...,N/1,2) Log(p(3,...,N/1,2)) \right) \quad (18)$$

As in the previous section, we can introduce a conditional entropy

$$\sigma(3,...,N/1,2) = -k_B \sum_{\varphi_3,...,\varphi_N} p(3,...,N/1,2) Log(p(3,...,N/1,2)) \quad (19)$$

Minimizing the free energy thus gives:

$$0 = U(1,2) + \sum_{i \neq 1,2} \sum_{\varphi_i} p(i/1,2) U(1,i) + \sum_{j \neq 1,2} \sum_{\varphi_j} p(i/1,2) U(i,2)$$
$$+ k_B T + k_B T Log(p(1,2)) - T\sigma(3,...,N/1,2) + A \quad (20)$$

The Lagrange parameter $A$ expresses the normalization condition of the probabilities.

In order to go further we need to evaluate $\sigma(3,...,N/1,2)$. To do so we first make the additivity assumption

$$\sigma(3,...,N/1,2) \cong \sum_{i \neq 1,2} \sigma(i/1,2) = -k_B \sum_{i \neq 1,2} \sum_{\varphi_i} p(i/1,2) Log(p(i/1,2)) \quad (21)$$

It is similar to what we had made in the previous section for simplifying $\sigma(2,...,N/1)$. Equ.(20) is then changed into

$$Log(p(1,2)) + \frac{U(1,2)}{k_B T} = -\sum_{i \neq 1,2} \sum_{\varphi_i} p(i/1,2) \frac{U(1,i)}{k_B T} - \sum_{j \neq 1,2} \sum_{\varphi_j} p(i/1,2) \frac{U(i,2)}{k_B T}$$
$$- \sum_{i \neq 1,2} \sum_{\varphi_i} p(i/1,2) Log(p(i/1,2)) - A \quad (22)$$

We then make the following superposition assumption

$$p(3)p(3/1,2) \cong p(3/1)p(3/2) \quad (23)$$

It is a more general formulation of Kirkwood's superposition approximation. Equ.(22) then becomes

$$Log(p(1,2)) + \frac{U(1,2)}{k_B T} = -\sum_{i \neq 1,2} \sum_{\varphi_i} \frac{p(1,i)p(i,2)}{p(i)} \left( Log(p(1,i)) + \frac{U(1,i)}{k_B T} \right)$$
$$-\sum_{j \neq 1,2} \sum_{\varphi_i} \frac{p(1,i)p(i,2)}{p(i)} \left( Log(p(i,2)) + \frac{U(i,2)}{k_B T} \right) - \sum_{i \neq 1,2} \sum_{\varphi_i} p(i/1,2) Log(p(i)) - A \quad (24)$$

We add to the right side of the above equation the null term $2\sum_{i\neq 1,2}\sum_{\varphi_i} p(i/1,2) - 2(N-2)$ so that equ.(24) is changed into

$$Log(p(1,2)) + \frac{U(1,2)}{k_BT} = \sum_{i\neq 1,2}\sum_{\varphi_i}\frac{p(1,i)p(i,2)}{p(i)}\left(1 - Log(p(1,i)) - \frac{U(1,i)}{k_BT}\right)$$
$$+ \sum_{j\neq 1,2}\sum_{\varphi_j}\frac{p(1,i)p(i,2)}{p(i)}\left(1 - Log(p(i,2)) - \frac{U(i,2)}{k_BT}\right) - \sum_{i\neq 1,2}\sum_{\varphi_i} p(i/1,2)Log(p(i)) + A' \quad (25)$$

### 4. Case of a classical fluid

Let us now consider a simple classical fluid; the system is made of N identical molecules in a volume V; N and V are arbitrary large, but the density N/V has a given finite value. An important quantity is the pair correlation function $g(1,2)$ defined by $p(1,2) = g(1,2)p(1)p(2)$. If the interaction terms depend only upon the relative position of the molecules, the probabilities are replaced by probability densities and the summations over the particle states by integrals over the positions; we can take the origin in 1. In the absence of any external field and assuming the isotropy of the potential, the one-particle density is uniform in the volume; as a consequence, the last summation in equ.(25) is a constant term which can be embedded in the normalization constant $A'$.

We put $\mathbf{R}_{12} = \mathbf{R}$, $\mathbf{R}_{13} = \mathbf{r}$, $\mathbf{R}_{32} = \mathbf{R} - \mathbf{r}$, $|\mathbf{R}_{12}| = R$, $|\mathbf{R}_{13}| = r$, and introduce $h(r) = g(r) - 1$ and $g^{(0)}(R) = \left(\frac{-U(R)}{k_BT}\right)$. Starting from equ.(25) and after some manipulations the pair correlation function is found to be the solution of

$$Log(g/g^{(0)})(R) \cong + \frac{N}{V}\int_{(V)} d^3\mathbf{r}\ h(|\mathbf{R}-\mathbf{r}|)\left(h - gLog(g/g^{(0)})\right)(r) + A'' \quad (26)$$

In order to go from equ.(25) to equ.(26) we have expressed $g(r)$ as $h(r) + 1$, which evidences a few constant terms that can be embedded in the normalization constant $A''$; $h(r)$ and $gLog(g/g^{(0)})(r)$ are integrable functions which can be convoluted. Besides, with respect to equ.(25) an additional factor 1/2 has been introduced in order to take into account the indiscernibility of the molecules.

Equ.(26) can be re-written as

$$g(R) \cong Cg^{(0)}(R)exp\left(\frac{N}{V}\int_{(V)} d^3\mathbf{r}\ h(|\mathbf{R}-\mathbf{r}|)\left(h - gLog(g/g^{(0)})\right)(r)\right) \quad (27)$$

The normalization factor C can be evaluated as follows. Let $v$ be a typical order of magnitude of the molecular volume; the normalization condition $\sum_{\varphi_j} p(j/1) = 1$ involves

$$\int_{(V)} \frac{d^3\mathbf{r}}{V} g(r) = 1 + o(v/V) \quad (28)$$

By combining equ.(27) and equ.(28) we have $C = 1 + o(v/V)$, hence $C \cong 1$. We finally obtain

$$Log\left(g/g^{(0)}\right)(R) \cong \frac{N}{V}\int_{(V)} d^3\mathbf{r}\ h(|\mathbf{R}-\mathbf{r}|)\left(h - gLog(g/g^{(0)})\right)(r) \quad (29a)$$

$$g(R) \cong g^{(0)}(R)exp\left(\frac{N}{V}\int_{(V)} d^3\mathbf{r}\ h(|\mathbf{R}-\mathbf{r}|)\left(h - gLog(g/g^{(0)})\right)(r)\right) \quad (29b)$$

Or in a more compact way

$$Log\left(\frac{g}{g^{(0)}}\right) = Nh \otimes \left(h - gLog(g/g^{(0)})\right) \quad (30a)$$

$$g \cong g^{(0)} exp\left(Nh \otimes \left(h - gLog(g/g^{(0)})\right)\right) \quad (30b)$$

So g(R) is the solution of an integral, non-linear equation which can hopefully be solved by iterations or by perturbation techniques.

If in the right hand side g(r) is replaced by its asymptotic value 1 equ.(29a) becomes

$$Log\left(g/g^{(0)}\right)(R) \cong \frac{N}{V}\int_{(V)} d^3\mathbf{r}\ h(|\mathbf{R}-\mathbf{r}|)\left(h - Log(g/g^{(0)})\right)(r) \quad (31a)$$

The ratio $\left(g/g^{(0)}\right)(r) = \left(g + k_B TLog(U)\right)(r)$ is often noted "$y(r)$" in the literature. Equ.(31a) can thus be written

$$Log(y)(R) \cong \frac{N}{V}\int_{(V)} d^3\mathbf{r}\ h(|\mathbf{R}-\mathbf{r}|)\left(h - Log(y)\right)(r) \quad (31b)$$

The above equation is precisely what results from the hypernetted chain approximation. In addition, by writing

$$Log(g/g^{(0)}) = Log\left(1 + \frac{g - g^{(0)}}{g^{(0)}}\right) \cong \frac{g - g^{(0)}}{g^{(0)}} = \frac{g}{g^{(0)}} - 1 \qquad (32)$$

i.e. $Log(y) \cong y - 1$, then $h - Log(y) \cong fy$ where $f = g^{(0)} - 1$ is Mayer's function. Equ.(29a) gives back the expression resulting from the Percus-Yevick approximation

$$y(R) \cong 1 + \left(\frac{N}{V} \int_{(V)} d^3\mathbf{r}\; h(|\mathbf{R} - \mathbf{r}|) f y(r)\right) \qquad (33)$$

## 5. Conclusion

We have presented an alternative derivation of the pair correlation function $g(R)$ for simple classical fluids. We have obtained an integral, non-linear equation which formally allows computing $g(R)$. That equation could be solved by iterative or perturbative method. It admits the equation resulting from the hyper netted chain approximation (and the Percus-Yevick approximation) as a limit case. It should be envisaged in further steps to try to numerically solve the equation, and then to extend the approach to more realistic intermolecular potentials.